\documentclass[aoas,MSNbibl,nameyear,dvips]{arximspdf}
\usepackage{graphicx}
\doi{10.1214/14-AOAS754} %kopijuoti is PTS
\volume{8}
\issue{3}
\pubyear{2014}
\firstpage{1728}
\lastpage{1749}
\docsubty{FLA}
\setattribute{copyright}{owner}{In the Public Domain}

\makeatletter

\newcommand{\IW}{\mathrm{IW}}
\newcommand{\N}{\mathrm{N}}
\newcommand{\G}{\mathrm{Ga}}
\newcommand{\E}{\mathrm{E}}
\newcommand{\cov}{\operatorname{cov}}
\newcommand{\bX}{\mathbf{X}}
\newcommand{\bA}{\mathbf{A}}
\newcommand{\bB}{\mathbf{B}}
\newcommand{\bS}{\mathbf{S}}
\newcommand{\bV}{\mathbf{V}}
\newcommand{\bT}{\mathbf{T}}
\newcommand{\bVhat}{\hat{\mathbf{V}}}
\newcommand{\bZ}{\mathbf{Z}}
\newcommand{\bgamma}{\bolds{\gamma}}
\newcommand{\bgammahat}{\hat{\bolds{\gamma}}}
\newcommand{\btheta}{\bolds{\theta}}
\newcommand{\bthetabar}{\bar{\bolds{\theta}}}
\newcommand{\bpsi}{\bolds{\psi}}
\newcommand{\bmu}{\bolds{\mu}}
\newcommand{\bbeta}{\bolds{\beta}}
\newcommand{\bbetahat}{\hat{\bolds{\beta}}}
\makeatother

\begin{document}
\begin{frontmatter}

\title{Modeling the effect of temperature on ozone-related
mortality\thanksref{T1}}
\runtitle{Effect of temperature on ozone-related mortality}

\begin{aug}
\author[A]{\fnms{Ander}~\snm{Wilson}\thanksref{T2,m1}\ead[label=e1]{ander\_wilson@ncsu.edu}},
\author[B]{\fnms{Ana~G.}~\snm{Rappold}\thanksref{m2}\ead[label=e2]{rappold.ana@epa.gov}},
\author[B]{\fnms{Lucas~M.}~\snm{Neas}\thanksref{m2}\ead[label=e3]{neas.lucas@epa.gov}}
\and
\author[A]{\fnms{Brian~J.}~\snm{Reich}\corref{}\thanksref{T3,m1}\ead[label=e4]{brian\_reich@ncsu.edu}}
\runauthor{Wilson, Rappold, Neas and Reich}
\affiliation{North Carolina State University\thanksmark{m1}
% US Environmental Protection Agency,
% US Environmental Protection Agency
and
US Environmental Protection~Agency\thanksmark{m2}}
\address[A]{A. Wilson\\
B.~J.~Reich\\
Department of Statistics\\
North Carolina State University\\
2311 Stinson Drive\\
Raleigh, North Carolina 27695\\
USA\\
\printead{e1}\\
\phantom{E-mail: }\printead*{e4}}
\address[B]{A.~G.~Rappold\\
L.~M.~Neas\\
Environmental Public Health Division\\
US Environmental Protection Agency\\
Chapel Hill, North Carolina 27599\\
USA\\
\printead{e2}\\
\phantom{E-mail: }\printead*{e3}}
\end{aug}
\thankstext{T1}{The research described in this article has been
reviewed by the National Health and Environmental Effects Research Laboratory,
US Environmental Protection Agency, and approved for publication.
Approval does not signify
that the contents necessarily reflect the views and policies of the
Agency, nor does the mention of trade
names of commercial products constitute endorsement or recommendation
for use.}
\thankstext{T2}{Supported in part by an appointment to the Research
Participation Program
for the US Environmental Protection Agency, Office of Research and Development,
administered by the Oak Ridge Institute for Science and Education
through an interagency
agreement between the US Department of Energy and EPA.}
\thankstext{T3}{Supported in part by the EPA (R835228) and NIH
(5R01ES014843-02).}

% HISTORY:
\received{\smonth{12} \syear{2013}}
\revised{\smonth{5} \syear{2014}}

% ABSTRACT
%
\begin{abstract}
Climate change is expected to alter the distribution of ambient ozone
levels and temperatures which, in turn, may impact public health. Much
research has focused on the effect of short-term ozone exposures on
mortality and morbidity while controlling for temperature as a
confounder, but less is known about the joint effects of ozone and
temperature. The extent of the health effects of changing ozone levels
and temperatures will depend on whether these effects are additive or
synergistic. In this paper we propose a spatial, semi-parametric model
to estimate the joint ozone-temperature risk surfaces in 95 US urban
areas. Our methodology restricts the ozone-temperature risk surfaces to
be monotone in ozone and allows for both nonadditive and nonlinear
effects of ozone and temperature. We use data from the
National \mbox{Mortality} and Morbidity Air Pollution Study (NMMAPS) and show that the
proposed model fits the data better than additive linear and nonlinear
models. We then examine the synergistic effect of ozone and temperature
both nationally and locally and find evidence of a nonlinear ozone
effect and an ozone-temperature interaction at higher temperatures and
ozone concentrations.
\end{abstract}

% KEYWORDS
\end{frontmatter}

\setcounter{footnote}{3}

The US Environmental Protection Agency (EPA) has concluded that current
scientific evidence supports a ``causal relationship'' between ozone
and respiratory health effects and a ``likely to be causal''
relationship between ozone and cardiovascular health effects and
mortality [\citet{EPA2013}]. Extreme temperatures, especially heat
waves, have also shown adverse associations with respiratory and
cardiovascular health [\citet{Bhaskaran2009,Turner2012}]. As a
photochemical air pollutant, ozone and temperature are both driven by
solar radiation and their association is enhanced by the temperature
dependence of other ozone precursors [\citet{Bloomer2009}]. Both
ambient temperatures and ground-level ozone are expected to increase in
the near future [\citet{EPA2009}] in response to anticipated climate
changes [\citet{IPCC2007Science}]. Under these circumstances, a better
understanding of the joint effects of temperature and ozone on human
health is essential for public health.

Most multi-city, time-series studies of ozone have focused on a main
ozone effect while controlling for potential confounding variables in a
generalized additive model. The highly influential analysis of
\citet{Bell2004} estimated a 0.25\% (0.12\%--0.39\%) increase in mortality
associated with a 10 ppb increase in same-day mean ozone at the
national level while controlling for confounders such as temperature
and weather conditions. Recent studies have found evidence that the
joint effects of ozone and temperature may not be additive [\citet
{Bell2006,Smith2009,Chen2013}]. Future climate change scenarios predict
increases in the high ends of the distributions of both the ozone and
temperature. Under these future conditions, the disease burden of high
ozone and temperature days may be different if the joint
ozone-temperature effect is synergistic instead of additive.

Estimating the existence and nature of an interactive effect is
nontrivial as temperature and ozone are both correlated with health
outcomes and with each other. In multi-city, time-series studies the
analysis is further complicated by the different ozone and temperature
ranges observed in each city, which makes pooling estimates across
cities challenging. Several studies of ozone-related daily mortality
have used a stratified model to examine the potential differences in
ozone effect by temperature and found a larger average ozone effect on
high temperature days compared to moderate temperature days [\citet
{Bell2008,Ren2008,Smith2009}]. However, average ozone levels are higher
on high temperature days (see Figure~\ref{figCityOzoneFig}, e.g.,
and Figure~1 in the supplementary material for additional details [\citet
{Wilson2014AOASsuppl}]) and the ozone effect may be larger at higher
ozone concentrations [\citet{Smith2009}]. This makes it unclear if the
larger average ozone effect on high temperature days is due to the
higher ozone on those days or an ozone-temperature interaction.

%
%f1 #&#
\begin{figure}

\includegraphics{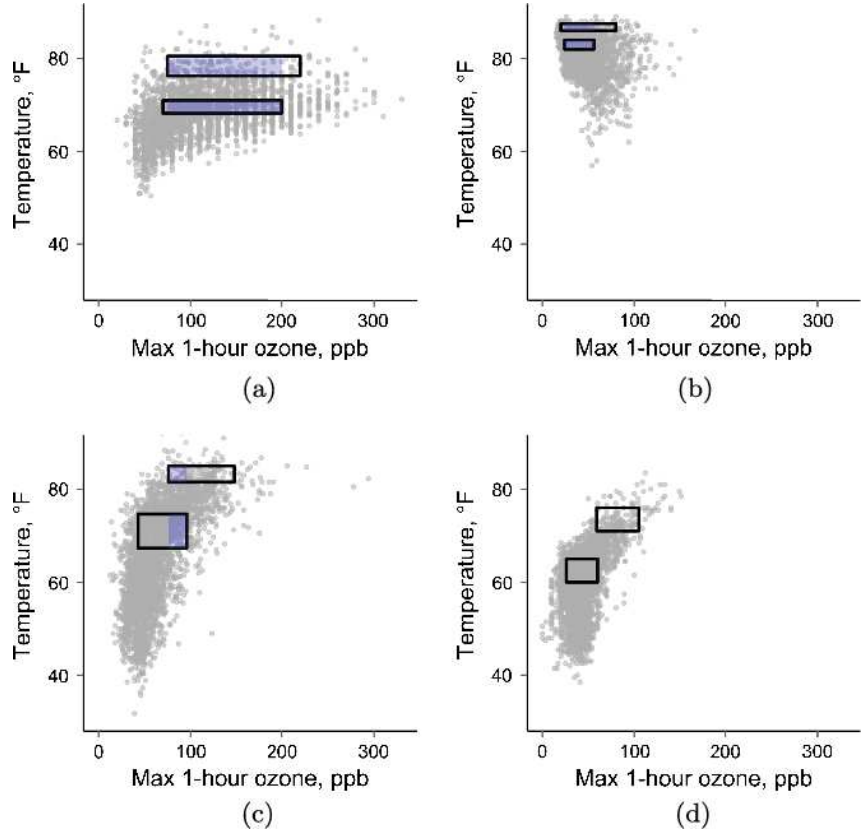}

\caption{Ozone-temperature distribution for selected cities. The upper
and lower boxes contain data for high temperature days and moderate
temperature days, respectively, over the observed ozone range. The
purple subsections highlight the common ozone range, the intersection
of the ozone ranges for high and moderate temperature days. Details on
the definition used to identify these days are in Section~\protect\ref
{sresultsinter}. Seattle is the only city in the data set that does not
have a common ozone range.
\textup{(a)}~Los Angeles,
\textup{(b)}~Miami,
\textup{(c)}~New York,
\textup{(d)}~Seattle.}\label{figCityOzoneFig}
\end{figure}
To estimate interaction, the concentration-response gradient for ozone
must be evaluated at different temperature ranges for the same ozone
range. Matching on ozone isolates interaction from the potential
confounding caused by higher ozone levels at higher temperatures and
nonlinear ozone effects. One way to do this is to estimate the full
ozone-temperature risk surface and evaluate the rate of change of the
risk surface in the ozone direction at different temperatures for a
constant ozone value. The gradient of the risk surface with respect to
ozone describes the ozone effect for each temperature and ozone
concentration. A changing gradient as a function of temperature for a
fixed ozone value signifies interaction between ozone and temperature.
We will refer to the gradient in the ozone direction as the log
relative risk (log RR) of ozone throughout, which approximates excess RR.

Recent studies have estimated the ozone-temperature risk surface in
single-city and independently in multi-city analyses [\citet
{Chen2013,Ren2008}]. The risk surfaces appear to be nonlinear and
nonadditive, but inference was limited to visual inspection. Instead,
these studies resorted to a stratified model with different linear
effects in each temperature stratum for a formal analysis instead of
making quantitative inference from the risk surfaces. In addition, the
multi-city approach used by \citet{Ren2008} estimates the risk surfaces
independently in each city, not allowing for sharing of information
between cities. The resulting risk surface estimates are highly
variable and do not allow for combining city-specific surfaces to
estimate a national risk surface.

In this paper, we propose a spatial monotone surface model to estimate
the ozone-temperature risk and log RR surfaces in multi-city
time-series studies. We model the ozone-temperature risk surface with
the outer-product of Bernstein polynomial basis functions and restrict
the surfaces to be monotone in ozone. The Bernstein polynomial
formulation enables closed-form representation of the risk surfaces and
log RR surfaces and facilitates comparisons between cites. It also
allows for sharing strength between nearby cities with a spatial model
for the basis coefficients.

We also present a two-stage approach to this model that reduces its
computational burden for large data [following the framework of
\citet{Dominici2002}]. In the two-stage approach, we introduce a
transformation that allows a different local basis expansion to be used
in each city that adapts to the city-specific ozone and temperature
ranges. This allows for the first-stage surface estimates to be
estimated with a basis expansion specifically tailored to the
ozone-temperature distribution in each city. However, the second stage
surfaces use a common basis expansion spanning the national
ozone-temperature distribution.

We use the proposed spatial monotone surface model to estimate the
national and city-specific log RR surfaces in 95 US urban areas using
data from the National Morbidity and Mortality Air Pollution Study
[NMMAPS; \citeauthor{Samet2000NMMAPS1} (\citeyear{Samet2000NMMAPS1,Samet2000NMMAPS2})]. Our results show
evidence of a synergistic ozone-temperature effect. At higher
temperatures and ozone values, the log RR of ozone tends to increase
with temperature. We find the ozone-temperature interaction increases
estimates of excess mortality at higher temperatures.

%%%%----------------------------------------------------------------------------------
%%%%----------------------------------------------------------------------------------
%s1 #&#
\section{Spatial monotone surface model}\label{model}
We assume the mortality count $Y_{ct}$ in city $c$ at time $t$ is
Poisson with log mean
%
%
%e1 #&#
\begin{equation}
\log\E(Y_{ct}) = f_c(\mathrm{ozone}_{ct}, \mathrm{temp}_{ct}) + g_c(\mathrm{confounders}_{ct}),
\label{stage1}
\end{equation}
where $f_c$ is the city-specific ozone-temperature risk surface and
$g_c$ controls for potential confounding variables. The model for daily
mortality is adopted from \citet{Bell2004} by moving daily mean
temperature from the confounder model $g_c$ to the risk model $f_c$ and
relaxing the assumptions on $f_c$ to include interaction and nonlinear
effects. The bivariate risk surface $f_c$ is modeled with a spatial
prior that imposes monotonicity in the ozone direction. The confounder
model $g_c$ includes linear and nonlinear functions of potential
confounders. In the subsections below we specify models for the
components of (\ref{stage1}).

%s1.1 #&#
\subsection{Ozone-temperature surface model}
We model $f_c$ as the outer-product of Bernstein polynomial basis
expansions of ozone and temperature [\citet{Lorentz1986,Tenbusch1997}].
This allows for a flexible regression surface including nonlinear and
nonadditive effects, but includes additive linear or polynomial effects
as special cases.
When making inference on a potentially nonlinear and nonadditive risk
function the log of the expected change in risk at each concentration
of ozone and temperature is expressed by the derivative of the log risk
surface $f_c$. Both the Bernstein polynomial regression function (the
log risk surface $f_c$) and its derivatives (the log RR surface) can be
expressed in closed form, facilitating analysis of the log RR of ozone.

The $k$th Bernstein polynomial basis function of order $M$ is
$b_k(x,M)={M \choose k} x^k(1-x)^{M-k}$ for $x\in[0,1]$. Denote ozone
as $x_1$, temperature as $x_2$ and $\mathbf{x}=(x_1,x_2)^T$. To scale the
data to the unit interval, we define the basis function
$B_{l,k}(x_l,M)=b_k[(x_l-\min_{ct} x_{l,ct})/r_l,M]$, where $r_l=\max
_{ct} x_{l,ct}-\min_{ct} x_{l,ct}$ and $l=1,2$ indicates ozone and
temperature, respectively. For notational simplicity let
$B_{j,k}(\mathbf
{x},M_1,M_2) = B_{1,j}(x_1, M_1)\times B_{2,k}(x_2, M_2)$. The
bivariate regression function~is
%
%
%e2 #&#
\begin{equation}
f_c(\mathbf{x}) = \sum_{j=0}^{M_1}
\sum_{k=0}^{M_2} \psi_{j,k,c}
B_{j,k}(\mathbf{x},M_1,M_2), \label{bpf}
\end{equation}
where $j$ and $k$ index the ozone and temperature basis expansions,
respectively. The first derivative of (\ref{bpf}) with respect to
ozone is
%
%
%e3 #&#
\begin{equation}
\frac{\partial f_c(\mathbf{x})}{\partial x_1} = M_1\sum_{j=0}^{M_1-1}
\sum_{k=0}^{M_2} (\psi_{j+1,k,c}-\psi
_{j,k,c} )B_{j,k}(\mathbf{x},M_1-1,M_2),
\label{deriv}
\end{equation}
the log RR of ozone. This is the change in log expected mortality
associated with a small increase in ozone.

For simplicity, we write the $(M_1+1)(M_2+1)$-vector of unknown
coefficients as
%
%
%e4 #&#
\begin{equation}
\bpsi_c= \pmatrix{ \psi_{0,0,c}, \ldots,\psi_{M_1,0,c},\psi_{0,1,c},\ldots,\psi_{M_1,M_2,c}}^{T}.
\end{equation}
Also, denote the $n_c\times(M_1+1)(M_2+1)$ basis expansion of ozone
and temperature in city $c$ over days $t=1,\ldots,n_c$ as
%
%
%e5 #&#
\begin{equation}
\bB(\mathbf{X}_c )= \bigl[B_{0,0}(
\bX_c, M_1,M_2),\ldots,B_{M_1,M_2}(\bX
_c, M_1,M_2) \bigr]^T.
\end{equation}

%s1.2 #&#
\subsection{Hierarchical model for monotonicity and spatial smoothing}\label{MSSsection}
We evaluate the bivariate association of current day's ambient
temperature and ozone with mortality based on the prior hypothesis of a
monotonic, but nonlinear, effect of ozone and a nonlinear effect of
temperature. The adverse health effects of short-term exposures to
ozone have been well established and described as causal to respiratory
effects and likely to be causal to mortality [\citeauthor{EPA2006} (\citeyear{EPA2006,EPA2013})].
Previous studies have estimated a monotone concentration-response
function using the same data without imposing specific shape
restrictions [\citet{Bell2006,Smith2009}]. A monotone relationship has
also been found in observational [\citet{Korrick1998,Ostro1993}] and
controlled exposure [\citet{Horstman1990,McDonnell2012}] human studies
of pulmonary function. Temperature effect has been described to have
``inverted J'' or ``U-shaped'' risk function [\citet
{Curriero2002}]. To
reflect this prior knowledge, we constrain the ozone effect to be
monotone, but leave the temperature effect unconstrained.

Bernstein polynomials are well suited for shape-restricted regression
[see \citet{Chang2007,Curtis2011,WangGhosh2012}]. From (\ref{deriv}),
a~sufficient condition for monotonicity in the ozone direction is $\psi
_{j+1,k,c}\ge\psi_{j,k,c}$ for all $j$ and $k$. We reparameterize the
coefficients as $\theta_{0,k,c}=\psi_{0,k,c}$ and $\theta
_{j,k,c}=\psi_{j+1,k,c}-\psi_{j,k,c}$ for $j>0$ using the matrix
%
%
%e6 #&#
\begin{equation}
\mathbf{T} = \mathbf{I}_{M_2+1}\otimes\pmatrix{ 1 & 0 & 0 & \cdots& 0 &
0
\cr
-1 & 1 & 0 & \cdots& 0 & 0
\cr
0 & -1 & 1 & \cdots& 0 & 0
\cr
\vdots&
\vdots& \vdots& \ddots& \vdots& \vdots
\cr
0 & 0 & 0 & \cdots& -1 &
1}_{(M_1+1)\times(M_1+1)}.
\end{equation}
With this parameterization $\btheta_c=\mathbf{T}\bpsi_c$ and
$f(\mathbf
{x};\btheta_c)$ is monotone in $x_1$ if $\theta_{j,k}\ge0$ for $j>0$.

The ozone-temperature surface can vary across cities. To borrow
strength across nearby cities while ensuring a monotone risk surface,
we model the basis coefficients with a truncated multivariate Gaussian
process (GP). To do this, we define the latent vector $\btheta^*_c$
and let $\theta_{0,k,c}=\theta^*_{0,k,c}$ and $\theta_{j,k,c}=\max
(0,\theta^*_{j,k,c})$ if $j>0$ for all $k$.

The prior on $\btheta^*_c$ is a multivariate Gaussian process with
mean $\E(\btheta^*_c )=\bmu$ and separable covariance function
%
%
%e7 #&#
\begin{equation}
\operatorname{cov} \bigl(\btheta^*_c,\btheta^*_{c'}
\bigr)=\exp\biggl[-\frac{d(c,c')}{\rho} \biggr] \times\mathbf{S}_2
\otimes\mathbf{S}_1,
\end{equation}
where $\bS_1$ is a $(M_1+1)\times(M_1+1)$ matrix capturing covariance
in the ozone direction, $\bS_2$ is a $(M_2+1)\times(M_2+1)$ matrix
capturing the covariance in the temperature direction, and the
exponential function captures spatial dependence. The distance function
$d(c,c')$ is the great circle distance between the centers of cities
$c$ and $c'$ in kilometers. In applications where the spatial units are
all neighboring, a conditional autoregressive model [CAR; \citet
{Banerjee2004,Gelfand2010,Lawson2013}] may be a reasonable alternative
to the distance-based GP. However, there are very few neighboring
cities with observed data in the NMMAPS data analyzed in Section~\ref{analysis}, so we chose the distance-based GP over the neighbor-based
CAR. The mean vector $\bmu$ has prior $\N(\mathbf{1}\mu_0,
\tau
\mathbf{S}_2\otimes\mathbf{S}_1 )$. While each component of the
separable covariance is not identifiable on its own, the product is
identifiable.

In many two-stage normal--normal models, $\bmu$ is often
interpreted as the estimate of the national average risk [e.g.,
\citet{Bell2004}]; however, in this model $\bmu$ represents the
mean of the
latent process $\btheta^*_c$ and not the process defining the shape
restricted surfaces of interest and thus does not have the same
interpretation. To obtain the estimate of the national average log RR
surface presented in Section~\ref{analysis}, we use the precision
weighted average of the city-specific log RR surfaces which are
realizations of a truncated process $\btheta_c$ defining the modeled
shape restricted surfaces. The precision weighted estimate gives more
weight in the tails to cities with data extending to those regions of
the surface, whereas the population weighted and unweighted cities do
not reflect the different ozone-temperature distribution in each city.

%s1.3 #&#
\subsection{Confounder model}\label{SectionConfounder}
We define $g_c$ as a generalized additive model that includes linear
and nonlinear effects for potential confounders. This is the confounder
model used by \citet{Bell2004}, with the same degrees of freedom,
\mbox{excluding} daily mean temperature which is now in $f_c$. The confounder
model includes an age-specific intercept ($<$65, 65--74, $\ge$75),
categorical variables for day of week, and smooth functions of time
interacted with age group with seven degrees of freedom per year
(natural splines). We control for additional effects of weather with
natural cubic spline of the 3-day running mean of temperature with six
degrees of freedom, natural cubic spline of dewpoint with 3 degrees of
freedom and of the 3-day running mean of dewpoint with 3 degrees of
freedom. The confounder model can be represented as the linear model
$g_c(\bZ_c)=\bZ_c\bgamma_c$. The prior for the confounder regression
coefficients is $\pi(\bgamma_c)\propto1$. While the prior for the
risk surface $f_c$ includes a spatial component, the prior on the
confounder model is independent between cities.

%s2 #&#
\section{A two-stage approach for large data sets}\label{twostage}
For large data sets, estimating the ozone-temperature risk surfaces is
computationally intensive. To ease computation, we break the model into
two stages, similar to the approach used by \citet{Dominici2002} and
\citet{Bell2004}. The two-stage approach approximates the spatial
monotone model presented in Section~\ref{model} but with reduced
computational burden. In the first stage, we estimate (\ref{stage1})
separately in each city with no monotonicity constraint or spatial
smoothing using computationally efficient quasi-likelihood estimation.
Then, we use the first-stage parameter estimates as data for a Bayesian
hierarchical model that spatially smooths the risk surfaces and
constrains each city's risk surface to be monotone in ozone. The
results given in Section~\ref{analysis} are produced using this
two-stage approach.

A natural approach for the city-specific estimates is to use the same
basis expansion in each city. However, the observed ozone and
temperature ranges vary dramatically from city to city. As a result,
some basis functions are well supported at the national level but not
supported in some cities individually. This can yield unstable
first-stage estimates. To extract as much information as possible from
each city in the first stage, we use a local basis expansion in each
city that spans only the observed ozone and temperature ranges in that
city. In the second stage a common basis expansion is used for all
cities to approximate the full model described in Section~\ref{model}.
Figure~2 in the supplementary material shows the first and second stage basis expansions
for six cities [\citet{Wilson2014AOASsuppl}].

%s2.1 #&#
\subsection{Stage 1: City-specific GLM regression}
The first-stage basis expansions are scaled to the city-specific ozone
and temperature ranges. In city $c$ the first-stage basis\vspace*{1pt} functions are
$b^c_{l,k}(x_l,M^c)=b_k[(x_l-\min_{t} x_{l,ct})/r_l^c,M^c]$, where
$r_l^c=\max_{t} x_{l,ct}-\min_t x_{l,ct}$, for $l=1,2$. The
first-stage estimates can also vary in the order of the basis
expansion, allowing for smaller $M_1^c$ and $M_2^c$ in cities with
smaller ozone and temperature ranges, respectively. Using this basis
expansion, the first-stage model of $f_c$ is
%
%
%e8 #&#
\begin{equation}
f_c(\mathbf{x}) = \sum_{j=0}^{M^c_1}
\sum_{k=0}^{M^c_2} \beta_{c,j,k}
b^c_{1,j}\bigl(x_1, M^c_1
\bigr)b^c_{2,k}\bigl(x_2, M^c_2
\bigr), \label{bp}
\end{equation}
with unknown first-stage parameters $\beta_{c,j,k}$, $j=0,\ldots,M_1^c$
and $k=0,\ldots,M_2^c$. The confounder model remains as
specified in Section~\ref{SectionConfounder}. Denote the first-stage
\mbox{estimate} as $(\bbetahat{}_c^T,\bgammahat{}_c^T)^T$ and $\cov
[(\bbetahat{}_c^T,\bgammahat{}_c^T)^T ] = \bV_c$, where $\bV_c$
can be partitioned as
%
%
%e9 #&#
\begin{equation}
\bV_c = \pmatrix{ \bV_{c,11} & \bV_{c,12}
\cr
\bV_{c,21} & \bV_{c,22} },
\end{equation}
and $\mathbf{b}_c(\bX_c)$ is the $n_c\times(M^c_1+1)(M^c_2+1)$
matrix of
first-stage basis expansions. We assume that $n_c>(M^c_1+1)(M^c_2+1)$
in all cities.

The first-stage parameters $\bbeta_c$ correspond to different basis
functions than those used in other cities and in the global expansion
in (\ref{bpf}). Hence, $\bbeta_c$ are not directly comparable across
cities or with $\bpsi_c$. This difference is resolved in the second
stage (Section~\ref{SectionStage2}).

%s2.2 #&#
\subsection{Stage 2: Bayesian model for stage 1 output}\label{SectionStage2}
In the second stage, we reparameterize the first-stage risk surface
estimates in terms of the global basis expansion (\ref{bpf}). This
provides a common set of parameters $\btheta_c$ to estimate $f_c$,
$c=1,\ldots,n$, with our spatial monotone model described in
Section~\ref{MSSsection}. Setting the first-stage parameterization of
the risk function $f_c(\bX_c)=\mathbf{b}_c(\bX_c)\bbeta_c$ equal\vspace*{1pt} to the
second-stage parameterization $f_c(\bX_c)=\bB(\bX_c)\bT^{-1}\btheta_c$
and solving for $\bbeta_c$ yields $\bbeta_c=\bA_c\btheta_c$, where
%
%
%e10 #&#
\begin{equation}
\bA_c= \bigl[\mathbf{b}_c^T(
\bX_c)\mathbf{b}_c(\bX_c)
\bigr]^{-1}\mathbf{b}_c^T(\bX_c)
\bB(\bX_c)\bT^{-1}.\label{Amatrix}
\end{equation}
The quantity $\bA_c\btheta_c$ is the projection of the second-stage
log risk surface onto the column space of the first-stage basis expansion.

Most two-stage approaches use the likelihood $\bbetahat_c\sim\mathrm
{N}(\bbeta_c,\bVhat_{c,11})$ in the second stage, where $\bbetahat
_c$ and $\bVhat_c$ are the first-stage estimates of $\bbeta_c$ and
$\bV_c$ [e.g., \citet{Dominici2002,Bell2004}]. For our model
$\bbeta
_c$ does not have the same meaning in each city. We instead replace
$\bbeta_c$ with $\bA_c\btheta_c$ and use the second-stage likelihood
%
%
%e11 #&#
\begin{eqnarray}
\pmatrix{ \bbetahat_c
\cr
\bgammahat_c } \Big|
\btheta_{c}, \bgamma_c, \bVhat_c &\sim& \N\left[
\pmatrix{ \bA_c\btheta_c
\cr
\bgamma_c },
\bVhat_c\right].\label{stage1likelihood}
\end{eqnarray}
To complete the second-stage model, we put a flat prior on $\bgamma$
and use the prior model described in Section~\ref{MSSsection} for
$\btheta^*$ (and thus for $\btheta$).

%s2.3 #&#
\subsection{Priors and computational details}
We estimate (\ref{stage1}) with quasi-li\-kelihood methods and the \texttt{glm} function in \texttt{R} to obtain the first-stage parameter estimates
$\bbetahat_c$, $\bgammahat_c$ and $\bVhat_c$, for $c=1,\ldots,N$,
where $\bVhat_c$ is the covariance matrix based on the Fisher's
information matrix returned from the \texttt{glm} package. We use an
offset proportional to the log population to account for different
population sizes. To complete the Bayesian specification of the
second-stage model, we use the hyperpriors:
%
%
%e12 #&#
\begin{eqnarray}
\mu_0 &\sim& \N\bigl(0,\tau_0^2 \bigr),\nonumber
\\
\mathbf{S}_2&\sim&\IW(M_2+2,\mathbf{I} ),\nonumber
\\
\mathbf{S}_1&\sim&\IW(M_1+2,\mathbf{I} ),
\\
\tau&\sim&\G(a_\tau,b_\tau)\quad\mbox{and}
\nonumber
\\
\log(\rho) &\sim& \N\bigl(\mu_\rho,\sigma^2_\rho
\bigr).\nonumber
\end{eqnarray}

We are interested in the posterior of $(\btheta{}_c^T,\bgamma{}_c^T)^T$.
To expedite computation, we perform MCMC sampling on the marginal
posterior of $\btheta_c$, marginalizing over $\bgamma_c$ which is
immediate from (\ref{stage1likelihood}).

The parameters $\bmu$, $\mu_0$, $\tau$, $\mathbf{S}_1$ and $\mathbf{S}_2$
have simple conjugate forms and are updated with Gibbs sampling. The
latent $\btheta^*_c$ are sampled with a Gibbs sampler using a mixture
of truncated normals. The range does not have a closed-form full
conditional. We sample $\log(\rho)$ with a random-walk
Metropolis--Hastings sampler. The full conditional for all parameters,
acceptance ratio for $\rho$ and MCMC algorithm are provided in supplementary material,
Appendix D [\citet{Wilson2014AOASsuppl}].

To get the predicted values used for cross-validation in Section~\ref{CVsection}, we need the posterior mean of $\bgamma_c$,
%
%
%e13 #&#
\begin{equation}
\E(\bgamma_c|\btheta_c,\bbetahat_c,
\bgammahat_c,\bV_c ) = \bgammahat_c+
\bV_{c,21}\bV_{c,11}^{-1} (\bA_c\bthetabar
_c-\bbetahat_c ),
\end{equation}
where $\bthetabar_c$ is the posterior mean of $\btheta_c$. Hence, no
MCMC is required to get the posterior mean of $\bgamma_c$, rather, it
can be computed in closed form using the marginal posterior estimate.

%%%%----------------------------------------------------------------------------------
%%%%----------------------------------------------------------------------------------
%%%%----------------------------------------------------------------------------------
%%%%----------------------------------------------------------------------------------

%s3 #&#
\section{Analysis of the ozone-temperature log RR surfaces}\label{analysis}
In this section we estimate the city-specific and national average log
RR surfaces using the two-stage approach presented in Section~\ref
{twostage} and examine the nature of the ozone effect at different
temperatures and ozone levels. We use the NMMAPS data and estimate the
risk surfaces using same-day 1-hour maximum ozone and mean
temperature, and estimate the surfaces for the same 95 US urban areas
used in \citet{Bell2004}. The NMMAPS data contains time-series
data for 1987 through 2000 with daily mortality counts by age group and daily
measurements of ozone, meteorological conditions and co-pollutants. We
limit the analysis to April through October, or ozone season.

%s3.1 #&#
\subsection{Cross-validation}\label{CVsection}
We performed cross-validation to determine if the spatial monotone risk
surface model fits the data better than alternative models and to
determine the order of polynomials to use. For each city we used 80\%
of the data as a training set and fit each model to those data. We
compared models with the deviance of the 20\% holdout sample using the
predicted values. The cross-validation deviance is $2[\hat
{Y}_{ct}-Y_{ct} \log(\hat{Y}_{ct})-\log(Y_{ct}!)]$, where $\hat
{Y}_{ct}$ is the predicted mortality count on holdout sample day $t$ in
city $c$.

The comparison models are as follows:
\begin{longlist}[(3)]
\item[(1)] A nonspatial monotone risk surface model that replaces the
multivariate Gaussian process on $\btheta^*_c$ with an independent
multivariate normal model at each site. Hence, $\btheta^*_c$ are
independent $\N(\bmu,\mathbf{S}_2\otimes\mathbf{S}_1)$.
\item[(2)] A spatial unconstrained risk surface version that models
$f(\mathbf
{x};\btheta_c)$ directly with a (nontruncated) multivariate Gaussian
process prior for $\btheta_c$, hence removing the monotonicity constraint.
\item[(3)] A nonspatial unconstrained model that combines the two previously
described simplifications of the monotone risk surface model.
\item[(4)] The NMMAPS model presented in \citet{Bell2004} with a linear
ozone effect and natural spline for temperature.
\item[(5)] An additive model that replaces the linear ozone effect in the\break
NMMAPS model with an unconstrained spline.
\end{longlist}

For the risk surface models we use the noninformative hyperparameters
$\tau_0=100$, $a_\tau=0.001$ and $b_\tau=0.001$. For the range
parameter we let $\mu_\rho=7$ and $\sigma_\rho=10$ to provide a
diffuse prior centered around 1000 km, which is about the range
estimate reported in \citet{Smith2009}. For the additive model we use
the same priors formulation as the nonspatial, unconstrained model, but
only use a basis expansion of ozone in $f_c$ since temperature is
accounted for in the confounder model. The second-stage model was run
for 500,000 iterations and the first half were discarded as burn-in. We
thinned the posterior sample, keeping every 50th draw due to high
autocorrelation, primarily in the range parameter.
Figures~4 and 5 in the supplementary material show trace plots of the posterior sample used for analysis
[\citet{Wilson2014AOASsuppl}].

%
%
%t1 #&#
\begin{table}[b]
\tabcolsep=0pt
\caption{Difference in cross-validation deviance from the linear additive model (NMMAPS model)}\label{tablecv}
\begin{tabular*}{\tablewidth}{@{\extracolsep{\fill}}@{}lcccc@{}}
\hline
& &\multicolumn{3}{c@{}}{\textbf{Above 95th percentile}}\\[-6pt]
& &\multicolumn{3}{c@{}}{\hrulefill}\\
& \textbf{Overall} & \textbf{Ozone} & \textbf{Temperature} & \textbf{Both}\\
\hline
Spatial monotone & $-$473.2 & $-$42.3 & $-$38.7 & $-$21.1 \\
Nonspatial monotone & $-$459.9 & $-$40.2 & $-$38.5 & $-$20.7 \\
Spatial unconstrained &$-$461.4 & $-$38.0 & $-$35.7 & $-$18.9 \\
Nonspatial unconstrained & $-$447.9 & $-$38.0 & $-$39.5 & $-$21.4
\\[3pt]
Additive ($M_1=4$) & $-$353.9 & $-$24.6 & $-$28.7 & $-$16.5 \\
\hline
\end{tabular*}
\end{table}

Table~\ref{tablecv} shows the difference in cross-validation deviance
from the\break NMMAPS model for the other five models. For the spatial
monotone model the best performing model had second-stage expansion of
order $M_1=7$ and $M_2=9$ and first-stage city-specific expansion
orders $M_1^c=\max(r^c_1M_1/r_1,6)$ and $M_2^c=\max(r^c_2M_2/r_2,4)$.
Hence, the first-stage expansions are smaller and proportional to the
ratio of the city-specific range ($r^c_l$) to the national range
($r_l$) down to a minimum order. CV results for this order are
presented in Table~\ref{tablecv}. Additional results for other order
expansions are provided in Appendix B in the supplementary material [\citet{Wilson2014AOASsuppl}].

Overall, the spatial monotone model has the smallest cross-validation
deviance. All four risk surface models have deviances well below those
of the linear and nonlinear additive models. Hence, the data support
using a nonadditive, nonlinear model. About one-third of the difference
between the spatial monotone and spatial unconstrained risk surface
models is in the high ozone area where there is less data and the
signal can be weak. The monotonicity constraint helps inform the model
and reduces variance in the tails while making a relatively smaller
difference over the areas with richer data. In addition, the
monotonicity constraint smooths the risk surfaces because there cannot
be mini peaks and valleys in the surface. This too helps improve fit in
the holdout data set and it seems reasonable to assume the exposure
response relationship should be fairly smooth. Because we are
interested in the tail behavior, we compare the deviance on holdout
sample days with ozone above each city's 95th percentile, days with
temperature above the 95th percentile and the intersection of the two.
The spatial monotone model had the lowest deviance for higher ozone,
but the nonspatial unconstrained model fit slightly better for high
temperature and the intersection. We proceed to analyze the data with
the spatial monotone model with $M_1=7$ and $M_2=9$.

%s3.2 #&#
\subsection{Analysis of the national average log RR surface}\label{sectionnationalRR}
\begin{figure}

\includegraphics{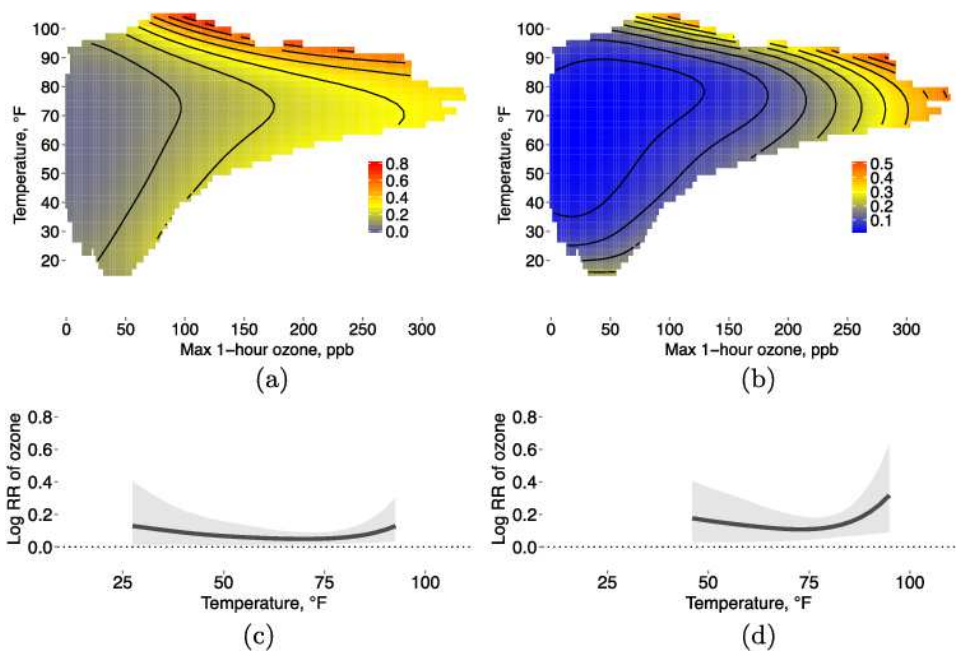}

\caption{Panels \textup{(a)} and \textup{(b)} show the pointwise mean and standard
deviation of the national log RR surfaces. Panels \textup{(c)} and \textup{(d)} show a
cross section of the log RR surface with ozone fixed at 50 and 100 ppb,
respectively, along with 95\% posterior intervals. Log RR is in percent
change in mortality per one ppb increase in ozone.}\label{fignationalRR}\label{fignationalRR50}\label{fignationalRR100}
\end{figure}

The national ozone-temperature log RR surface (the derivative of $f$
with respect to ozone) shows an association between daily 1-hour
maximum ozone and mortality with posterior probability greater than
0.99. The ozone effect is greater at higher ozone concentrations and at
higher temperatures. Hence, a one ppb increase in ozone is associated
with a larger increase in mortality on high temperature days and days
with higher ozone levels. Figure~\ref{fignationalRR}(a) shows the
national log RR surface plotted over the range of observed data. This
is the pointwise average over the city-specific log RR surfaces
weighted by the pointwise city-specific precision. Figure~\ref{fignationalRR50}(c)~and~(d) show the log RR of ozone
at 50 ppb and 100 ppb, respectively, as functions of temperature, along
with 95\% posterior intervals. All results for log RR are the gradient~(\ref{deriv}) multiplied by 1000. This can be interpreted as the
expected percent change in mortality associated with a 10 ppb increase
in ozone, where 10 ppb was chosen to be consistent with other publications.

%
%f2 #&#

\begin{figure}

\includegraphics{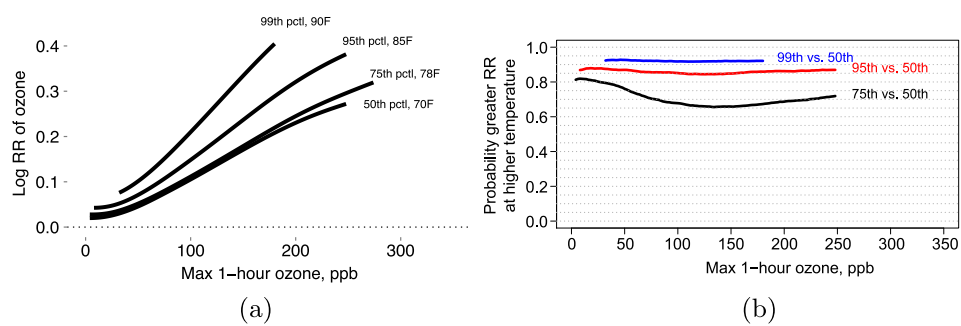}

\caption{Comparison of the log RR at the 50th, 75th, 95th and 99th
percentiles of temperature. The estimates are plotted over the range of
ozone values observed at that temperature in at least 5 cities.
\textup{(a)}~Shows the mean log RR for each cross section and
\textup{(b)}~shows the pointwise posterior probability that log RR is greater at the high temperature. Log RR is
in percent change in mortality per one ppb increase in ozone.
\textup{(a)}~Mean log RR at four percentiles of temperature (50th, 75th, 95th, 99th),
\textup{(b)}~posterior probability that RR is greater at higher temperatures than median
temperatures for each ozone value.}\label{figeyelash}\label{figeyelashpr}
\end{figure}

At higher ozone concentrations, temperature has a larger modifying
effect. Figure~\ref{figeyelash} shows the log RR at the 50th, 75th,
95th, and 99th percentiles of temperature as a function of ozone. For
each temperature, log RR increases at higher ozone concentrations. As
temperature increases from the median, the log RR increases
monotonically for all ozone values, however, the difference between the
cross-sections at low ozone values is very small. The posterior
probability that RR is greater at the 99th percentile than the 50th
percentile is about 0.92 [Figure~\ref{figeyelashpr}(b)].
%
%
%f3 #&#

\begin{figure}[t]

\includegraphics{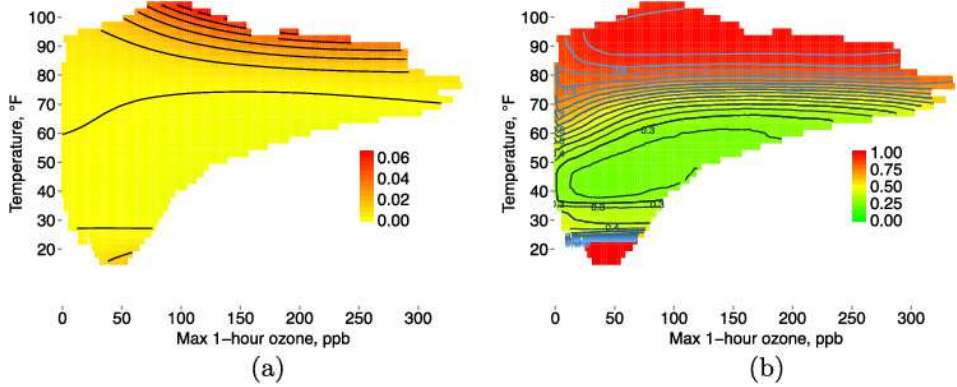}

\caption{National interaction surface and pointwise posterior
probability of positive interaction.
\textup{(a)}~Shows the national interaction surface which is the cross-derivative of the
log risk surface or the derivative of the log RR surface with respect
to temperature. This shows how log RR changes with temperature and
quantifies the interaction at each point.
\textup{(b)}~Shows the posterior probability that the national interaction surface is greater than 0.}\label{figNatInt}
\end{figure}

The cross-derivative of the risk surface provides a more complete
picture of a departure from additivity. The cross-derivative surface is
%
%
%e14 #&#
\begin{equation}\label{crossderiv}
\frac{\partial^2 f_c(\mathbf{x})}{\partial
x_1\,\partial x_2} = M_1M_2\sum
_{j=0}^{M_1-1}\sum_{k=0}^{M_2-1}
(\theta_{j,k+1,c}-\theta_{j,k,c} )B_{j,k}(
\mathbf{x},M_1-1,M_2-1).\hspace*{-25pt}
\end{equation}
We refer to (\ref{crossderiv}) as the interaction surface, as it is
the rate of change in the log RR surface with respect to temperature.
With an additive model the cross-derivative is zero. Figure~\ref{figNatInt} shows the national interaction surface and the pointwise
posterior probability that the interaction is greater than zero. The
national interaction surface shows a synergistic effect at higher ozone
and temperatures. This synergism occurs with a posterior probability
greater than 0.95 over the part of the surface with temperature equal
to 90$^{\circ}$F. At lower temperatures, the mean interaction
is negative, although not with a high posterior probability in this
summer-only analysis.

%
%f4 #&#

%
%
%f5 #&#
\begin{figure}

\includegraphics{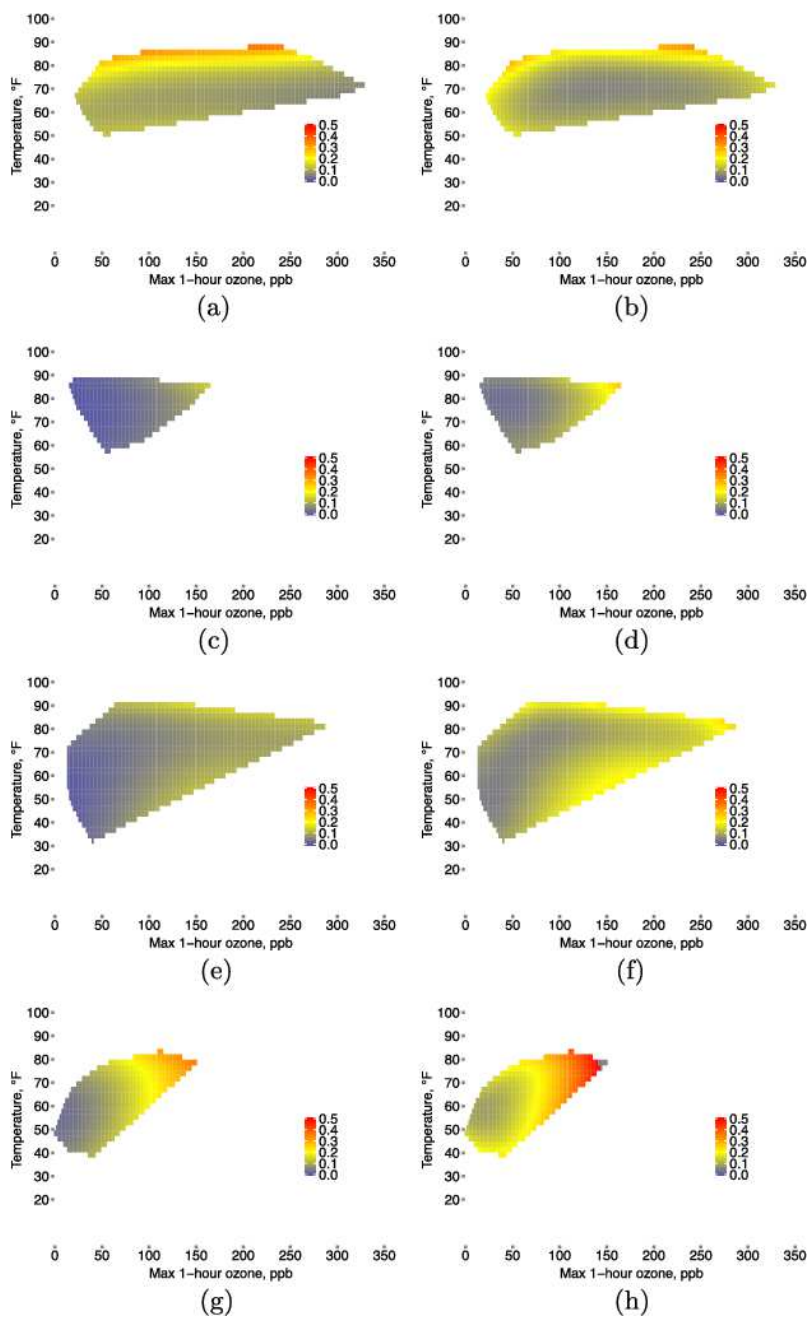}

\caption{Log RR surfaces for selected cities \textup{(left)} and their
pointwise standard deviations \textup{(right)}. Surfaces are plotted only over
the range of data observed data for that city. The cities were selected
for being geographically diverse and having varied ozone and
temperature ranges.
\textup{(a)}~Los Angeles log RR surface,
\textup{(b)}~Los Angeles SD,
\textup{(c)}~Miami log RR surface,
\textup{(d)}~Miami SD,
\textup{(e)}~New York log RR surface,
\textup{(f)}~New York SD,
\textup{(g)}~Seattle log RR surface,
\textup{(h)}~Seattle SD.}\label{figCitySurfaces}
\end{figure}

%s3.3 #&#
\subsection{Analysis of the city-specific log RR surfaces}\label{sresultsinter}
We now examine the city-specific surfaces and interaction at the city
level. Figure~\ref{figCitySurfaces} shows examples of four\
city-specific log RR surfaces and their pointwise standard deviations.
While the national log RR surface shows a high posterior probability of
a positive ozone effect, the city-specific ozone effect, averaged over
the observed days in each city, is positive with posterior probability
greater than 0.95 in six cities (Los Angeles, CA; Dallas/Ft. Worth, TX;
Houston, TX; San Jose, CA; Oakland, CA; and St. Louis, MO).

To evaluate the interaction effect, we compare high and moderate
temperature days in each city. We define high temperature days as those
with temperatures between the 95th and 99th city-specific temperature
percentiles and moderate temperature days to have temperatures between
the 50th and 75th percentiles. Within each range we include only days
between the 10th and 90th percentiles of ozone in order to minimize the
influence of days with extreme ozone values in either direction.
Figure~\ref{figCityOzoneFig} outlines these days in black for four
cities. By using city-specific percentiles this definition of high and
moderate temperature days adapts to each city's weather; however, like
previous studies that used a stratified model to compare log RR at
different temperatures, it does not account for the different ozone
distributions of high and moderate temperature days. To remove the
effect of different ozone ranges, we limit the high and moderate
temperature regions to a common ozone range, indicated by the purple
box in Figure~\ref{figCityOzoneFig}.

\begin{figure}

\includegraphics{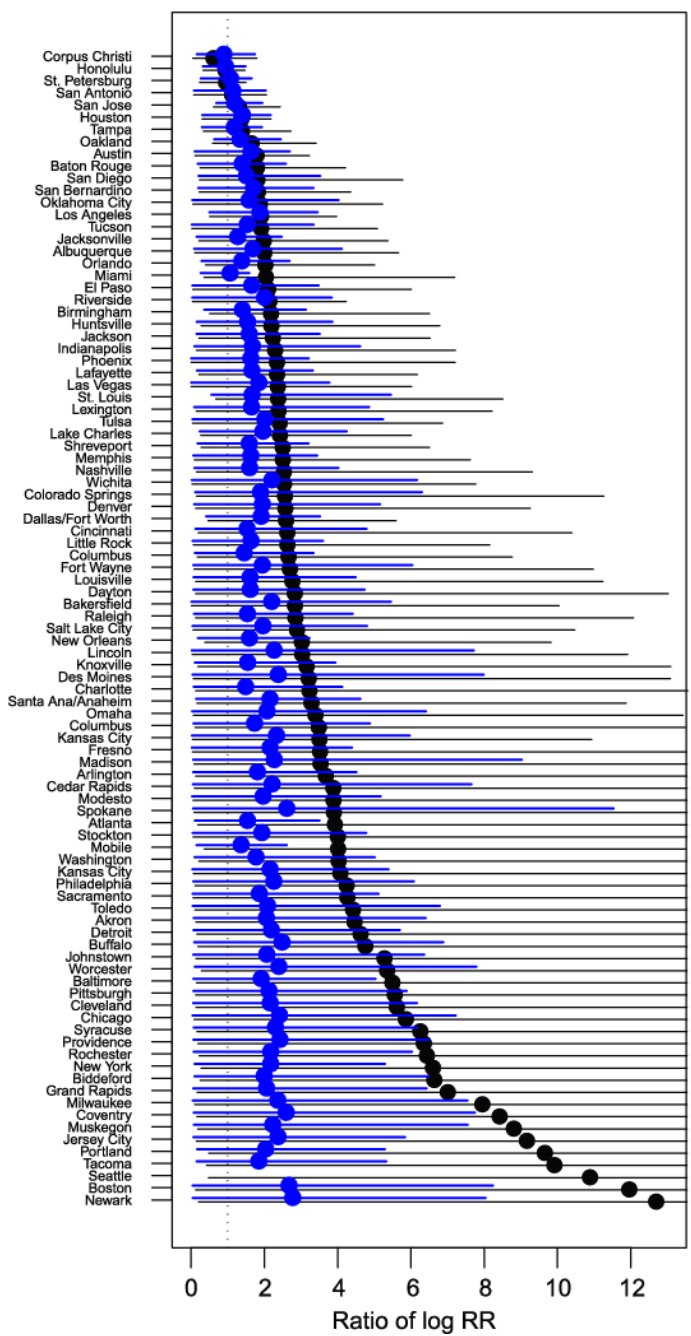}

\caption{Comparison of the log RR at high temperatures and moderate
temperatures as defined in Section~\protect\ref{sresultsinter}. The
ratio of log RR over the observed ozone range is shown in black and common ozone
range is in blue. The posterior mean and 95\% interval are shown.}\label{figIntCompare}
\end{figure}

\begin{figure}

\includegraphics{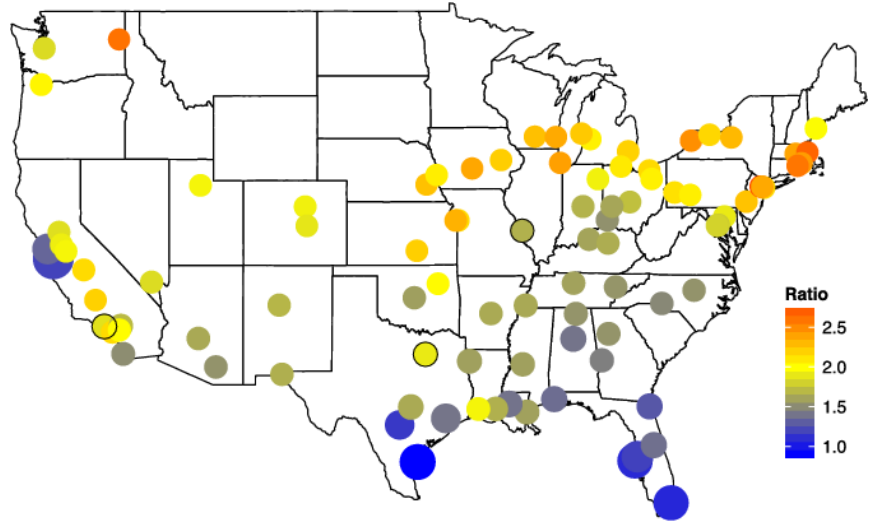}

\caption{Map of the ratio of log RR at high temperatures and moderate
temperatures over the common ozone range as defined in Section~\protect
\ref{sresultsinter}. The RR at higher temperatures is greater than the RR
at moderate temperatures with posterior probability at least 0.8 in the
cities outlined in black (Dallas/Ft Worth, Los Angeles, and St.
Louis).}\label{figIntMap}
\end{figure}

On high temperature days the log RR is larger than on moderate
temperature days over the observed ozone range in most cities, but the
difference is greatly reduced when comparing only over the common ozone
range. Figure~\ref{figIntCompare} compares the ratio of mean log RR on
high temperature days to mean log RR on moderate temperature days for
both ozone ranges. The large reduction in the ratio when limiting to a
common ozone range suggests that much of the difference in log RR
between high and moderate temperature days is due to the higher ozone
levels on high temperature days in conjunction with the nonlinearity of
the ozone effect. Over the common ozone range, the ratio ranged from
about 1 to 2.5. Most of the cities with larger levels of interaction
over the common ozone range are in the north where there is a larger
difference between the temperatures on high and moderate temperature
days (Figure~\ref{figIntMap} and Table~\ref{tableinteraction}).
%
%
%f6 #&#
%
%
%f7 #&#
%
%
%t2 #&#
\begin{table}[b]
\tabcolsep=0pt
\caption{Mean log RR on high and moderate temperature days over the
observed and common ozone rates by region (in percent change in
mortality per 10 ppb increase in ozone)}\label{tableinteraction}
\begin{tabular*}{\tablewidth}{@{\extracolsep{\fill}}@{}lcccc@{}}
\hline
&\multicolumn{2}{c}{\textbf{Observed ozone range}} & \multicolumn{2}{c@{}}{\textbf{Common ozone range}}\\[-6pt]
&\multicolumn{2}{c}{\hrulefill} & \multicolumn{2}{c@{}}{\hrulefill}\\
&\multicolumn{1}{c}{\textbf{High temp.}} & \multicolumn{1}{c}{\textbf{Moderate temp.}} & \multicolumn{1}{c}{\textbf{High temp.}} & \multicolumn{1}{c@{}}{\textbf{Moderate temp.}}\\
\hline
Indust. midwest & 0.25~(0.10) & 0.17~(0.06) & 0.24~(0.09) & 0.19~(0.07) \\
Northeast & 0.12~(0.09) & 0.06~(0.05) & 0.11~(0.09) & 0.07~(0.05) \\
Northwest & 0.14~(0.06) & 0.07~(0.03) & 0.12~(0.05) & 0.09~(0.04) \\
Southern CA & 0.13~(0.07) & 0.08~(0.05) & 0.11~(0.06) & 0.09~(0.06) \\
Southeast & 0.17~(0.39) & 0.14~(0.33) & 0.17~(0.39) & 0.15~(0.33) \\
Southwest & 0.14~(0.08) & 0.08~(0.05) & 0.13~(0.08) & 0.08~(0.05) \\
Upper midwest & 0.10~(0.07) & 0.06~(0.04) & 0.09~(0.07) & 0.06~(0.04)
\\[3pt]
National & 0.15~(0.04) & 0.09~(0.02) & 0.13~(0.04) & 0.10~(0.03) \\
\hline
\end{tabular*}
\tabnotetext[]{}{Note: The regions from the NMMAPS data are used [see \citeauthor{Samet2000NMMAPS1} (\citeyear{Samet2000NMMAPS1,Samet2000NMMAPS2})].}
\end{table}

These results are similar to many previous studies that find
statistical significance at the national level but in only a handful of
cities [see \citet{Bell2004}, e.g.]. The motivation of
hierarchical modeling and sharing information between cities in our
study is to increase the power to estimate effects. While the posterior
probabilities in most cities do not strongly support interaction at the
city level, the city-specific point estimates imply interaction in many
cities, a result consistent with the national results presented in
Section~\ref{sectionnationalRR}.

%s3.4 #&#
\subsection{Analysis of excess mortality}
Overall we see a trend of a synergistic ozone-temperature effect, both
in the national log RR surface and interaction surface, and in the
city-specific estimates. This interaction at higher ozone levels and\vadjust{\goodbreak}
temperatures leads to a larger estimate of excess mortality. Table~\ref
{tablemodelcompare} compares the expected change in mortality
associated with an increase from median ozone and temperature to the
95th percentile of ozone and temperature for the additive linear model
(NMMAPS), the additive nonlinear model and the monotone, spatial risk
surface model by region. The general trends are the same, with the
largest effects observed in the industrial midwest, but the risk
surface estimates are more homogeneous across regions.
%
%
%t3 #&#
\begin{table}
\tabcolsep=0pt
\caption{Percent increase in mortality associated with an increase
from the medians of ozone and temperature to the 95th percentiles of
ozone and temperature using different models}\label{tablemodelcompare}
\begin{tabular*}{\tablewidth}{@{\extracolsep{\fill}}@{}lccc@{}}
\hline
&\multicolumn{1}{c}{\textbf{Additive linear}} & \multicolumn{1}{c}{\textbf{Additive nonlinear}} & \multicolumn{1}{c@{}}{\textbf{Surface}}\\
\hline
Indust. midwest & 4.57~(0.77) & 3.27~(1.65) & 4.13~(0.42) \\
Northeast & 5.61~(0.94) & 5.88~(1.92) & 5.31~(0.48) \\
Northwest &3.92~(0.82) & 0.90~(1.84) & 2.42~(0.58) \\
Southern CA & 2.77~(0.80) & 3.89~(2.57) & 3.88~(0.77) \\
Southeast & 0.70~(0.52) & 3.15~(1.20) & 3.23~(0.38) \\
Southwest & 2.89~(0.84) & 4.71~(2.04) & 4.49~(0.64) \\
Upper midwest & 1.33~(1.21) & 2.91~(2.08) & 4.82~(0.62)
\\[3pt]
National & 3.06~(0.30) & 3.54~(0.75) & 3.98~(0.24) \\
\hline
\end{tabular*}
\end{table}

%s4 #&#
\section{Discussion}\label{discussion}
In this paper we propose a two-stage procedure to estimate
city-specific ozone-temperature risk surfaces. To accommodate different
temperature and ozone ranges in different cities, we use local basis
expansions in the first stage. The first-stage results are combined in
the second-stage model using a global basis expansion and
spatially-varying coefficients to allow for different ozone-temperature
effects by city.

We evaluated the model fit with respect to the modeling assumptions
using cross-validation and the results indicated that monotonically
nondecreasing shape-restricted ozone effect in a spatial model was best
supported by the data. These results suggest that the monotonic
constraint helps inform the model where data is sparse while allowing
the constraints of linearity of the association to be relaxed. Previous
attempts to analyze the bivariate effect of temperature and ozone on
mortality have either over-smoothed with loss of information on
potential interaction or under-smoothed, resulting in biologically
implausible scenarios where increasing doses of ozone may alternate
between being beneficial and detrimental to health [\citet{Cheng2012,Burkart2013}, \citeauthor{Ren2008} (\citeyear{Ren2008,Ren2008b})].

Our analysis of the data in 95 US cities provides additional evidence
that the effects of ozone and temperature are nonlinear, and depart
from simple additivity between ozone and temperature. Specifically, the
national average log RR surface indicates that the RR of ozone is
higher on high temperature days and this interaction is most pronounced
in the northern US, where summer temperatures have the most variability.

The results of this study have important implications toward
understanding the nature of the joint effects of ozone and temperature
on mortality. To compare the ozone effect between different temperature
strata, it is important that the distributions of ozone be similar
within each strata. High correlation between temperature and ozone
violate this requirement and thus call for careful consideration and
interpretation of methods. Indeed, the results suggest that the higher
ozone estimates at high temperatures in stratified studies are
primarily due to the nonlinear effect of ozone coupled with higher
concentrations of ozone formed at high temperatures. However, the
results also suggest that modification of risk is present at higher
temperature levels.

\citet{IPCC2007Science} raised concerns about whether current air
quality management practices can adequately protect public health under
the future climate regimes. Climate projections show increases in
extreme ozone and temperature days over a significant portion of the US
[\citet{Hogrefe2004,Kunkel2007,Tagaris2007,Wu2008}]. The synergistic
ozone-temperature effect estimated in this paper implies that the
disease burden of these extreme weather days is greater than would be
estimated with additive models. This is despite the relatively small
size of the interaction effect compared to the nonlinear main effect at
higher ozone levels.

In our paper we have developed methodology to capture nonlinear
interactions between temperature and ozone effects. This methodology
could be used in other health effects analyses and potentially beyond.
For example, there is increasing interest in identifying joint effects
of multiple pollutants [\citet{NRC2004}]. Recent work on
multiple-pollutant modeling include \citet{Kalendra2010} and
\citet{Bobb2013}, who study the joint effect of ozone and fine particulate
matter. Another application is studying cumulative effects using
various lagged pollution or temperature variables [\citeauthor{Heaton2012} (\citeyear{Heaton2012,Heaton2013}), \citet{Schwartz2000,Welty2005}]. \citet{Bell2004} found
that the current day ozone exposures (lag 0), the same measure analyzed
in this paper, have the largest effect on total mortality and
cardiovascular and respiratory deaths. However, previous day exposures
(lags 1 and 2) were also significantly associated with daily mortality,
as was the cumulative effect of exposures of the previous week (lags
0--6). For any two predictors our method would apply directly; extending
the model to include a high-dimensional surface for the joint effect of
several predictors will be challenging. To accommodate several
predictors, our approach could be modified to have an additive
structure [\citet{Hastie1990}] with main effect curves for each
predictor and two-dimensional interaction surfaces for pairs of
variables thought to interact.

\begin{supplement} \label{supp}
\stitle{Appendices}
\slink[doi]{10.1214/14-AOAS754SUPP} %[doi,text={...}] - jei reikia suskaldyti doi
\sdatatype{.pdf}
\sfilename{aoas754\_supp.pdf}
\sdescription{Appendices referenced in the text are provided in the
supplementary appendix file [\citet{Wilson2014AOASsuppl}].
Appendix A: additional figures.
Appendix B: cross-validation results.
Appendix C: full conditional distributions.
Appendix D: MCMC algorithm.
Appendix E: trace plots.}
\end{supplement}


%suskaldyti doi

% imsref loaded by linak, 2014-07-17 13:56:28
% imsref loaded by linak, 2014-07-18 08:49:53
%
% imsref loaded by linak, 2014-07-22 13:31:21
% imsref loaded by linak, 2014-07-22 13:34:01

\printaddresses
\end{document}